\documentclass{ifacconf}

\usepackage{graphicx}      
\usepackage{natbib}        
\usepackage{amsmath,amsfonts,amssymb}
\usepackage{tikz}
\usepackage{subfigure}
\usepackage{paralist}
\usepackage{pgfplots}
\usetikzlibrary{arrows,shapes,trees}
\newtheorem{Definition}{Definition} 
\newtheorem{Remark}{Remark} 

\newcommand{\mc}[1]{\mathcal{#1}}

\newcommand{\ik}{^{(k)}}
\newcommand{\ikt}{^{(k)\top}}

\begin{document}
\begin{frontmatter}

\title{Cooperative Estimation for Synchronization of Heterogeneous Multi-Agent Systems Using Relative Information\thanksref{footnoteinfo}} 

\thanks[footnoteinfo]{This work was supported
by the German
Research Foundation (DFG) through
the Cluster of Excellence in Simulation Technology (EXC 310/1) at the
University of Stuttgart and
the Australian Research Council under Discovery Projects funding scheme (project DP120102152).}

\author[First]{Jingbo Wu}, 
\author[Second]{Valery Ugrinovskii}, 
\author[First]{Frank Allg\"ower}

\address[First]{Institute for Systems Theory and Automatic Control, University of Stuttgart,
   70550 Stuttgart, Germany (e-mail: \{jingbo.wu, allgower\}@ ist.uni-stuttgart.de)}
\address[Second]{
School of Information Technology and Electrical Engineering, University of New South Wales at ADFA 
Australian Defence Force Academy,
Canberra, ACT 2600, Australia (e-mail: v.ougrinovski@adfa.edu.au).
}

\begin{abstract}                
In this paper, we present a distributed estimation setup where local agents
estimate their states from relative measurements received from their
neighbours. In the case of heterogeneous multi-agent systems, where only
relative measurements are available, this is of high relevance. The
objective is to improve the scalability of the 
existing distributed estimation algorithms by restricting the agents to
estimating only their local states and those of immediate neighbours. The
presented estimation algorithm also guarantees robust performance against
model and measurement disturbances. It is shown that it can be integrated
into output synchronization algorithms. 
\end{abstract}


\end{frontmatter}
\section{INTRODUCTION}
Estimator design has been an essential part of controller design ever since
the development of state-space based controllers. A milestone was laid by
the Kalman Filter in 1960 \citep{Kalman1960}. 

While in the classical estimator design one estimator is used for one
system, designing distributed estimators have gained attention since a
distributed Kalman Filter was presented in \citep{OlfatiSaber:2005vm},
\citep{OlfatiSaber2007}, \citep{Carli2008}. In a distributed estimator setup, multiple
estimators create an estimate of the system's state, while cooperating
with each other. In this setup, even when every single estimator may be
able to obtain an estimate of the state on its own, cooperation reduces the
effects of model and measurement disturbances
\citep{SS-2009}.
Also, the situations are not uncommon where every single estimator is unable
to obtain an estimate of the state on its own and cooperation becomes an
essential prerequisite \citep{Ugrinovskii2011},
\citep{Ugrinovskii2011a}. The node estimators may even not have a model 
of the full system, but only know a part of the system \citep{Stankovic2009}. 

Apart from distributed estimation, distributed control and coordination of
multi-agent system have become an important area in systems control
research, with a wide range of possible applications such as vehicle
platooning, UAV coordination, and energy networks. Starting from the
consensus problem of integrators \citep{OlfatiSaber2007}, \citep{Ren2007}
the boundaries have been pushed further and further, extending the theory
to linear high-order systems \citep{Fax2004}, \citep{Tuna2008}, dynamic
controllers \citep{Scardovi2009}, heterogeneous systems
\citep{Wieland2009}, \citep{Wieland2011}, and many more. In these problems,
a distributed control scheme is typically applied, i.e. for each agent in
the network a separate \emph{local controller} is implemented that can only
receive information from the agent and its neighbors. Therefore, an
important aspect of this setup is distinguishing between absolute
information gained from individual local measurements, $y_k=h_k(x_k)$, and
relative information gained through the network, $y_k - y_j$.  
For instance, in the framework of synchronizing vehicle formations some
information, such as the vehicle position, can only be determined from relative
measurements, i.e., distances between the vehicles. Some fundamental properties for estimation based on relative information with static agents have been presented in \citep{Barooah2007}. The case of estimating homogeneous multi-agent systems with relative information was considered in \citep{Acikmese2011}.
For heterogeneous multi-agent systems, this problem becomes particularly
challenging. In \citep{Wu2012} a solution was presented that was developed
from the Internal Model Principle for Synchronization
\citep{Wieland2009}. However, it is subject to some geometric conditions
that restrict the class of systems that can be considered. In
\citep{Grip2012}, the same problem is considered and an observer-based synchronization method is presented that solves the problem for
leader-follower networks. 

In this paper, we tackle the synchronization problem of heterogeneous
multi-agent systems from the distributed estimation point-of-view. 
Our main contribution is the
development of a new framework for distributed
state estimation using relative measurements. In this new framework,
cooperation between the local estimators 
will be crucial due to lack of local detectability, therefore calling it
\emph{Cooperative Estimation}. These estimators allow us to estimate
absolute information purely based on relative information, similar to
\citep{Listmann2011}, but for a more general class of systems. Another purpose of the new framework
is to address scalability of the filter network. In multi-agent
coordination problems, scalability of the network is important, i.e. the
dimension of the local controllers should not increase with the size of the
network. However, direct applications of the existing algorithms such as those
reported in \citep{OlfatiSaber:2005vm}, \citep{Olfati-saber2006} and
\citep{Ugrinovskii2011}, \citep{Ugrinovskii2011a} result in the order of
the estimators growing with the size of the network. Our estimation
framework resolves this issue by restricting the agents to perform estimation
of the local states and the neighbour states only. In particular, we
present an $\mathcal{H}_\infty$ suboptimal design, which can handle
model and measurement disturbances. 

After designing the local
estimators, we analyze the synchronizing behavior of the controller
presented in \citep{Wieland2011} where these cooperating observers are employed.
It turns out that in addition to synchronization, we can show guaranteed
$\mathcal{H}_\infty$-performance for the closed loop system. 

The rest of the paper is organized as follows: In Section 2, we present
some mathematical preliminaries and the system class which is
considered. Section 3 presents the cooperative estimation design with
suboptimal $\mathcal{H}_\infty$-performance. The loop is closed in Section 4,
where we analyze the performance, when the estimates are used in an
synchronization setup. Section 5 depicts our results with a simulation
example. This paper is the extended \texttt{ArXiv}-version of \cite{Wu2014}.

\section{PRELIMINARIES}
\vspace{-0.05cm}
Throughout the paper the following notation is used: $I_n$ denotes the $n \times n$ identity matrix.  
Let $A$ be a quadratic matrix. If $A$ is positive definite, it is denoted $A > 0$, and $A < 0$, if $A$ is negative definite. The weighted norm of a vector $\|x\|_W$ is defined as $\sqrt{x^\top W x}$ and the unweighted norm $\|x\|$ is defined as $\sqrt{x^\top x}$, respectively.
$A\otimes B$ denotes the Kronecker product of two matrices $A$ and $B$.
 
 \vspace{-0.1cm}
\subsection{Communication graphs}
\vspace{-0.1cm}
In this section we briefly review some definitions on communication graphs.
We use a directed, unweighted graph $\mc{G} = (\mc{V} , \mc{E}, \mc{A})$ to describe the interconnections between the individual agents. 
$\mc{V} = \{v_1,...,v_N\}$ is the set of vertices, where $v_k \in \mc{V}$ represents the $k$-th agent. 
$\mc{E} \subset V \times V$ is the set of edges, which models information flow, i.e. the $k$-th agent receives information from the $j$-th agent if and only if $(v_j,v_k) \in \mc{E}$. $\mc{A}$ is the adjacency matrix, which encodes the edges, where $a_{kj} = 1$ if $(v_j,v_k) \in \mc{E}$ and $a_{kj} = 0$ otherwise.  
A \emph{path} from vertex $v_{i_1}$ to vertex $v_{i_l}$ is a sequence of vertices $\{ v_{i_1} , . . . , v_{i_l} \}$ such that $\{v_{i_j}, v_{i_{(j+1)}}\} \in \mc{E},j = 1,...,l-1$. Moreover, the incidence matrix is the $|\mathcal{E}| \times N$ matrix encoding the edges, where $E_{ij} = 1$, if vertex $j$ is the tail of edge $\mathcal{E}_i$ and $E_{ij} = -1$, if vertex $j$ is the head of edge $\mathcal{E}_i$.

\smallskip
\begin{Definition} 
A directed graph $\mc{G}$ is \emph{strongly connected}, if for every pair
of vertices $v_k$, $v_j$, $k,j=1,...,N, j \neq k$, there exists a path from
$v_k$ to $v_j$. 
\end{Definition}
Moreover, we use the definition of independent strongly connected components defined in \citep{Wieland2010b}.
\begin{Definition} 
An \emph{independent strongly connected component (iSCC)} of a directed graph $\mc{G}$ is a subgraph $\widetilde{\mc{G}}=(\widetilde{\mc{V}} , \widetilde{\mc{E}}, \widetilde{\mc{A}})$, which is strongly connected and satisfies $(v,\widetilde{v}) \not\in \mc{E}$ for any $v \in \mc{V} \backslash \widetilde{V}$ and $\widetilde{v} \in \widetilde{V}$.
\end{Definition}

\subsection{System model}
We consider a group of $N$ agents described by the differential equation
\begin{equation}\label{eq:LTI}
\begin{aligned}
\dot{x}_k &= A_k x_k + B_k u_k + \widetilde{B}_k \xi_k 
\end{aligned}
\end{equation}
where $x_k \in \mathbb{R}^{n_k}$ is the agents state variable, $u_k \in
\mathbb{R}^{m_k}$ is the agents control input, $\xi_k(t) \in
\mathcal{L}_2[0, \infty)$ is a $\mathcal{L}_2$-integrable disturbance
function. The agents are interconnected and the topology is represented by
a directed graph $\mc{G} = (\mc{V} , \mc{E}, \mc{A})$. It is assumed that
the output $y_k=C_k x_k \in \mathbb{R}^r$ is not available to the agents as controller
input. Instead, disturbed relative information of the
form   
\begin{equation*}
z_{kj}=y_j -y_k+\omega \eta_{kj} \in \mathbb{R}^r
\end{equation*}
is available to the controller at node $k$, if and only if there is an edge $(v_j, v_k)$, i.e. $a_{kj}=1$. Here, $\eta_{kj}(t) \in
\mathcal{L}_2[0, \infty)$ is also a $\mathcal{L}_2$-integrable disturbance
function and $\omega \in \mathbb{R}$ is a positive weight. 
Let $p_k$, $q_k$ denote the in-degree and out-degree of vertex $v_k$,
respectively. 
For all $k=1,...,N$, $z^{(k)\top} = \begin{bmatrix}
z_{kj_1^k}^\top & ... & z_{kj_{p_k}^k}^\top
\end{bmatrix}$ is defined as the vector of all available measurements at node $k$ and $\eta^{(k)\top} = \begin{bmatrix}
\eta_{kj_1^k}^\top & ... & \eta_{kj_{p_k}^k}^\top
\end{bmatrix}$ the corresponding measurement disturbance vector. Here, the set $\{j_1^k,...,j_{p_k}^k\}$ denotes the neighbors of agent $k$. However, for the sake of a simple notation, we will drop the superscript index $k$ and only write $\{j_1,...,j_{p_k}\}$ for the neighbors of agent $k$ in the following.

In this paper, we assume that the communication topology
of the network is identical to the measurement topology, i.e., agent $k$ is
able to receive information from agent $j$ through communication, if and
only if it measures the relative output $z_{kj}$.  

With the stacked vector $x = [x_1^\top,...,x_N^\top]^\top$, the global
system can be written as 

\begin{equation}\label{sys:global}
\begin{aligned}
\dot{x}=& \begin{bmatrix}
A_1 & & \\
& \ddots & \\
& & A_N
\end{bmatrix} \begin{bmatrix}
x_1 \\
\vdots \\
x_N
\end{bmatrix} + \begin{bmatrix}
B_1 & & \\
& \ddots & \\
& & B_N
\end{bmatrix}\begin{bmatrix}
u_1 \\
\vdots \\
u_N
\end{bmatrix} \\
&+
\begin{bmatrix}
\widetilde{B}_1 & & \\
& \ddots & \\
& & \widetilde{B}_N
\end{bmatrix}\begin{bmatrix}
\xi_1 \\
\vdots \\
\xi_N
\end{bmatrix}. \\
\end{aligned}
\end{equation}
The global output is the vector of stacked $z\ik$, which is
\begin{equation} \label{sys:C_global}
z =
\left( E \otimes I_r \right)
\begin{bmatrix}
C_1 & & \\
& \ddots & \\
& & C_N
\end{bmatrix} 
x + \omega \eta,
\end{equation}
where $E$ is the incidence matrix corresponding to the graph $\mc{G}$ and $\eta$ is the vector of stacked $\eta\ik$.

%


\section{Cooperative Estimation design}

For the sake of estimator design, in this section, it is assumed that $u_k=0$ for all $k=1,...,N$.

\subsection{Proposed estimation algorithm}

\begin{figure}\label{fig:estimator_setup}
\begin{center}
\setlength{\unitlength}{1mm}
\begin{tikzpicture}[scale=0.8, transform shape]
\node [circle, fill=gray!20] (a) at (0,2*0.87) {$\sum_1$};
\node [circle, fill=blue!20] (a_est) at (-3,2*0.87+1.5) {$\begin{bmatrix}
\hat x_1^{(1)} \\
\hat x_2^{(1)}
\end{bmatrix}$};
\node [circle, fill=gray!20] (b) at (2,2*0.87) {$\sum_2$};
\node [circle, fill=blue!20] (b_est) at (3.5,2*0.87+1.5) {$\begin{bmatrix}
\hat x_2^{(2)} \\
\hat x_3^{(2)}
\end{bmatrix}$};
\node [circle, fill=gray!20] (d) at (1,0) {$\sum_4$};
\node [circle, fill=blue!20] (d_est) at (-0.5,-1.5) {$\begin{bmatrix}
\hat x_4^{(4)} \\
\hat x_1^{(4)}
\end{bmatrix}$};
\node [circle, fill=gray!20] (c) at (3,0) {$\sum_3$};
\node [circle, fill=blue!20] (c_est) at (6,-1.5) {$\begin{bmatrix}
\hat x_3^{(3)} \\
\hat x_4^{(3)}
\end{bmatrix}$};
\draw [thick, ->] (b) -- (a);
\draw [thick, ->] (c) -- (b);
\draw [thick, ->] (d) -- (c);
\draw [thick, ->] (a) -- (d);
\draw [thick, dotted, blue, ->] (a) -- (a_est)node [pos=0.7, right] {\; $y_2-y_1$};
\draw [thick, dotted, blue, ->] (b) -- (b_est)node [pos=0.3, right] {$y_3-y_2$};;
\draw [thick, dotted, blue, ->] (c) -- (c_est)node [pos=0.7, left] {$y_4-y_3$};;
\draw [thick, dotted, blue, ->] (d) -- (d_est)node [pos=0.7, right] {$y_1-y_4$};;
\draw [thick, dotted, blue, ->] (b_est) -- (a_est)node [pos=0.5, above] {$\hat x_2^{(2)}$};
\draw [thick, dotted, blue, ->] (c_est) -- (b_est)node [pos=0.5, right] {$\hat x_3^{(3)}$};
\draw [thick, dotted, blue, ->] (d_est) -- (c_est)node [pos=0.5, above] {$\hat x_4^{(4)}$};
\draw [thick, dotted, blue, ->] (a_est) -- (d_est)node [pos=0.5, right] {$\hat x_1^{(1)}$};
\end{tikzpicture}
\caption{Example for the estimator structure. The inner graph represents the agents and the measurement topology, the outer graph represents the estimators and the communication topology.
\vspace{0.2cm}}
\end{center}
\end{figure}
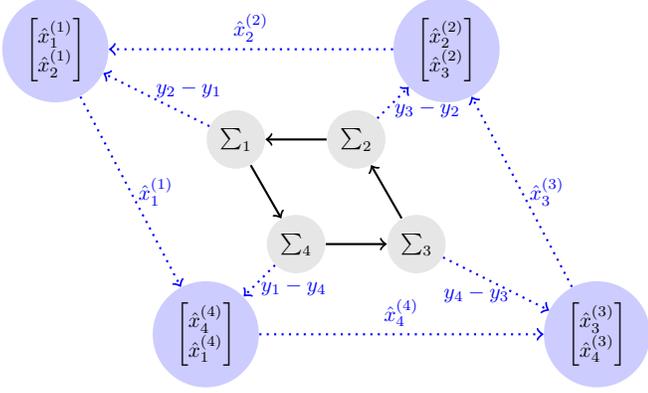

At every agent $k$, a local estimator is implemented that
estimates its own state $x_k$ and the states of its neighbors, i.e., $x_j$
with $a_{kj} = 1$. For this purpose, it is assumed that the estimator at
agent $k$ knows the models of its neighbors. The vector of local estimates
is defined as 
$\hat{x}^{(k)\top}= \begin{bmatrix}
\hat{x}^{(k)\top}_k &
\hat{x}^{(k)\top}_{j_1} &
\hdots &
\hat{x}^{(k)\top}_{j_{p_k}}
\end{bmatrix} \in \mathbb{R}^{\sigma_k}$
and the corresponding error vector is
\begin{equation*}
e^{(k)}= x^{(k)} -\hat{x}^{(k)} =\begin{bmatrix}
x_k - \hat{x}^{(k)}_k \\
x_{j_1} - \hat{x}^{(k)}_{j_1} \\
\vdots \\
x_{j_{p_k}} - \hat{x}^{(k)}_{j_{p_k}}
\end{bmatrix} \in \mathbb{R}^{\sigma_k},
\end{equation*}
where $\sigma_k$ is the dimension of the local estimator state. The estimator
architecture can be depicted as shown in Figure \ref{fig:estimator_setup}, with a cyclic topology as an example. 

The estimator dynamics are proposed as
\begin{equation}\label{eq:estimator}
\begin{aligned}
\dot{\hat{x}}^{(k)} =& A\ik \hat{x}^{(k)} + L\ik(z\ik-C\ik \hat{x}\ik) \\
& +K\ik \sum_{j=1}^N a_{kj}(M_j\ik \hat{x}_j^{(j)}- N_j\ik \hat{x}\ik)
\end{aligned}
\end{equation}
with initial condition $\hat{x}^{(k)}_0 = 0$. The filter gains to be designed are $L\ik \in \mathbb{R}^{\sigma_k \times rp_k}$ and $K\ik \in \mathbb{R}^{\sigma_k \times \sigma_k}$. $A\ik$ is the block-diagonal matrix with $A_k, A_{k_1},\ldots,A_{k_{p_k}}$ on its diagonal and $C\ik$ is the output matrix
\begin{equation*}
C\ik = \begin{bmatrix}
-C_k & C_{j_1} & \hdots & 0 \\
-C_k & 0 & \ddots & 0 \\
-C_k & 0 & 0 & C_{j_{p_k}}
\end{bmatrix}.
\end{equation*}

$M_j\ik \in \mathbb{R}^{\sigma_k \times n_j}$ is a matrix of the form
$ M_j^{(k)\top} = \begin{bmatrix}
0  &
I_{n_j}  &
0  
\end{bmatrix}$
and $N_j\ik \in \mathbb{R}^{\sigma_k \times \sigma_k}$ is a diagonal matrix such that 
\begin{equation*}
M_j\ik \hat{x}^{(j)}_j - N_j\ik \hat{x}\ik = \begin{bmatrix}
0  \\
\hat{x}^{(j)}_j - \hat{x}\ik_j \\
0  
\end{bmatrix}. 
\end{equation*}

We can now formulate the design problem of this section.

\textbf{Problem 1:}
Determine estimator gains $L\ik$, $K\ik$ in \eqref{eq:estimator} such that the
following two properties are satisfied simultaneously. 
\begin{compactenum}[(i)]
\item In the absence of model and measurement disturbances (i.e., when
  $\xi_k=0$, $\eta=0$), the estimation errors decay so that $e\ik \to 0$
  exponentially for all $k=1,...,N$. 

\item The estimators \eqref{eq:estimator} provide guaranteed
  $\mathcal{H}_{\infty}$ performance in the sense that 
\begin{equation}\label{eq:hinf-performance}
\begin{aligned}
&   \sum_{k=1}^N \int_0^\infty e^{(k)\top}W\ik e\ik  dt\\
&\leq \gamma^2 \sum_{k=1}^N \int_0^\infty (\| \xi\ik \|^2 + \| \eta\ik \|^2)dt + I_0,
\end{aligned}
\end{equation}
for $\xi^{(k)\top} = \begin{bmatrix}
\xi_k^\top & \xi_{j_1}^\top & \hdots & \xi_{j_{p_k}}^\top
\end{bmatrix}$
and a positive definite matrix $P\ik$. In (\ref{eq:hinf-performance}),
$W\ik$ is a positive semi-definite weighting matrix and $I_0=\sum_{k=1}^N
x^{(k)\top}_0 P\ik x\ik_0$ is the cost due to the observer's uncertainty about
the initial conditions of the agents. 
\end{compactenum}
The weighting matrix $W\ik$ is a design parameter and can be chosen as needed. As it will turn out in Chapter 4, a specific choice of $W\ik$ is needed in order to guarantee synchronization performance. 


\subsection{Filter gains design}
We define the matrices
\begin{align*}
Q\ik =& P\ik A + A^\top P\ik - G\ik C\ik - (G \ik C\ik )^\top \\ & - F\ik \sum_{j=1}^N a_{kj}N_j\ik - (F\ik \sum_{j=1}^N a_{kj}N_j\ik)^\top \\
& + \alpha P\ik + q_k \pi_k \begin{bmatrix}
 P\ik_{11} & 0 \\
 0 & 0
\end{bmatrix},\\
\widetilde{B}\ik &= \begin{bmatrix}
\widetilde{B}_k & 0 & 0 & 0 \\
0 & \widetilde{B}_{j_1} & \hdots & 0 \\
0 & 0 & \ddots & 0 \\
0 & 0 & 0 & \widetilde{B}_{j_{p_k}}
\end{bmatrix},
\end{align*}
where $P\ik \in \mathbb{R}^{\sigma_k \times \sigma_k}$ is a symmetric,
positive definite matrix and $P\ik_{11} \in \mathbb{R}^{n_k \times n_k}$ is
the top-left submatrix of $P\ik$. $\pi_k$ and $\alpha$ are positive 
constants which will later play the role of design parameters. With these
definitions, we are ready to present our main result. 

\begin{thm}
Consider a group of $N$ interconnected agents described by \eqref{eq:LTI},
\eqref{sys:global}, \eqref{sys:C_global}. Let a collection of matrices
$F\ik$, $G\ik$ and $P\ik$, $k=1,\ldots,N$, be a solution of the LMIs 
\begin{equation}\label{LMI}
\begin{aligned}
\!\!\begin{bmatrix}
Q\ik \!\! + \!\! W\ik & \!\!\! -\omega G\ik & P\ik \widetilde{B}\ik & \!\! \vline & F\ik M\ik_{j_1} & \hdots & F\ik \!\! M\ik_{j_{p_k}} \!\! \\
-\omega G^{(k)\top} & -\gamma^2 I & 0 & \!\! \vline & 0 & 0 & 0\\ 
(P\ik \widetilde{B}\ik)^\top & 0 & -\gamma^2 I & \!\! \vline & 0 & 0 & 0 \\
\hline & & & \!\! \vline \\
(F\ik \! M\ik_{j_1})^\top \!\!\! & 0 & 0 &  \!\! \vline & -\! \pi_{j_1} \! P_{11}^{(j_1)} & 0 & 0\\
\vdots & 0 & 0 & \!\! \vline & 0 & \ddots & 0 \\
(F\ik \! M\ik_{j_{p_k}})^\top \!\!\! & 0 & 0 & \!\! \vline & 0 & 0 & -\pi_{j_{p_k}} \!\! P_{11}^{(j_{p_k})} \!\!
\end{bmatrix} \\
< \!\! 0
\end{aligned}
\end{equation}
for all $k=1,...,N$, then Problem 1 admits a solution of the form
\begin{equation}\label{eq:filtergains}
\begin{aligned}
L\ik &= (P^{(k)})^{-1} G\ik \\
K\ik &= (P^{(k)})^{-1} F\ik.
\end{aligned}
\end{equation}
\end{thm}

\begin{pf}
The estimator error dynamics at node $k$ are
\begin{align*}
\dot{e}\ik =& (A\ik - L\ik C\ik )e\ik - \omega L\ik \eta\ik + \widetilde{B}\ik \xi\ik \\ &+ K\ik \sum_{j=1}^N a_{kj}(M_j\ik e^{(j)}_j- N_j\ik e\ik),
\end{align*}
 where $\xi^{(k)\top} = [\xi_k^\top, \xi_{j_1}^\top, \hdots, \xi_{j_{p_k}}^\top]$ and
 \begin{equation*}
 \widetilde{B}\ik = \begin{bmatrix}
 \widetilde{B}_k & 0 & 0 & 0 \\
 0 & \widetilde{B}_{j_1} & \hdots & 0 \\
 0 & 0 & \ddots & 0 \\
 0 & 0 & 0 & \widetilde{B}_{j_{p_k}}
 \end{bmatrix}.
 \end{equation*}

We use a Lyapunov function
\begin{equation*}
V(e)= \sum_{k=1}^N \underbrace{e^{(k)\top}P^{(k)}e\ik}_{V\ik(e\ik)},
\end{equation*}
where $V\ik(e\ik)$ are the individual components of $V(e)$.

The Lie derivative of $V\ik(e\ik)$ is
\begin{align*}
\dot{V}\ik(e\ik) =& 2 e^{(k)\top} P\ik (A\ik - L\ik C\ik) e\ik \\
&+ 2 e^{(k)\top} P\ik(- \omega L\ik \eta\ik + \widetilde{B}\ik \xi\ik) \\ 
&+ 2 e\ikt P\ik K\ik \sum_{j=1}^N a_{kj}(M_j\ik e^{(j)}_j - N_j\ik e\ik) \\
=& 2 e^{(k)\top} \! P\ik \!\!\! \left( \!\! A\ik \!\!-\!\! L\ik C\ik \!\!-\!\! K\ik \! \sum_{j=1}^N a_{kj}N_j\ik \!\! \right) \!\! e\ik \\ 
&+ 2 e^{(k)\top} P\ik (- \omega L\ik \eta\ik + \widetilde{B}\ik \xi\ik) \\
&+ 2 e\ikt P\ik K\ik \sum_{j=1}^N a_{kj}(M_j\ik e^{(j)}_j)
\end{align*}

With the filter gains \eqref{eq:filtergains}
and the LMIs \eqref{LMI} it can be obtained that
\begin{align*}
\dot{V}\ik(e) =& e^{(k)\top} Q\ik e\ik\\
&- e^{(k)\top}\left(\alpha P\ik + q_k \pi_k \begin{bmatrix}
 P\ik_{11} & 0 \\
 0 & 0
\end{bmatrix}\right)e\ik\\
&- 2 e^{(k)\top} \omega G\ik \eta\ik + 2 e^{(k)\top} P\ik \widetilde{B}\ik \xi\ik \\
&+ 2 e\ikt F\ik \sum_{j=1}^N a_{kj}(M_j\ik e^{(j)}_j) \\
\leq & \sum_{j=1}^N a_{kj} \pi_{j}e^{(j)T}_j P^{(j)}_{11}e^{(j)}_j - e^{(k)\top} W\ik e\ik \\ 
& +\gamma^2 \eta^{(k)\top}\eta\ik + \gamma^2 \xi^{(k)\top} \xi\ik \\
&- \alpha e^{(k)\top} P\ik e\ik - q_k \pi_k e^{(k)\top}_{k}  P\ik_{11}e^{(k)}_{k}.
\end{align*}
Summing up the $V\ik$s, it holds for $V$ that
\begin{equation*}
\begin{aligned}
\dot{V}(e) \leq & \underbrace{\sum_{k=1}^N \sum_{j=1}^N a_{kj} \pi_{j}e^{(j)T}_j P^{(j)}_{11}e^{(j)}_j}_{=\sum_{k=1}^N q_k \pi_{k}e^{(k)\top}_k P^{(k)}_{11}e^{(k)}_k} - \sum_{k=1}^N e^{(k)\top} W\ik e\ik\\ 
& + \sum_{k=1}^N\gamma^2 \eta^{(k)\top}\eta\ik + \sum_{k=1}^N\gamma^2 \xi^{(k)\top} \xi\ik \\
&- \sum_{k=1}^N \alpha e^{(k)\top} P\ik e\ik -  q_k \pi_k \sum_{k=1}^N e^{(k)\top}_{k} P\ik_{11}e^{(k)}_{k}
\end{aligned}
\end{equation*}

\begin{equation}\label{eq:Vdot}
\begin{aligned}
\dot{V}(e) \leq& -\alpha \sum_{j=1}^N  \underbrace{e^{(k)\top} P\ik e\ik}_{V\ik} - \sum_{k=1}^N e^{(k)\top} W\ik e\ik\\ 
& + \sum_{k=1}^N\gamma^2 \eta^{(k)\top}\eta\ik + \sum_{k=1}^N\gamma^2 \xi^{(k)\top} \xi\ik \\
\end{aligned}
\end{equation}
Integrating both sides of \eqref{eq:Vdot} on the interval $[0, T]$, we obtain
\begin{equation*}
\begin{aligned}
& V(e(T))+ \sum_{k=1}^N \int_0^T e^{(k)\top}W\ik e\ik dt \\
&\leq \gamma^2 \sum_{k=1}^N \int_0^T (\| \xi\ik \|^2 + \| \eta\ik \|^2)dt + \sum_{k=1}^N e^{(k)\top}_0 P\ik e\ik_0.
\end{aligned}
\end{equation*}
As $V(e(T))\geq 0$ and with the zero initial conditions of the observer states, it follows that
\begin{equation*}
 \sum_{k=1}^N \int_0^T e^{(k)\top}W\ik e\ik dt \leq \gamma^2  \sum_{k=1}^N \int_0^T (\| \xi\ik \|^2 + \| \eta\ik \|^2)dt + I_0.
\end{equation*}
Letting $T\to \infty$, this satisfies Property (ii) of Problem 1.

Moreover, if $\xi_k=0$ and $\eta_k=0$ for all $k=1,...,N$, then it
follows from (\ref{eq:Vdot})
 that
\begin{equation*}
\dot{V}(e) \leq -\alpha V,
\end{equation*}
which implies that Property (i) of Problem 1 holds.

\flushright $\blacksquare$
\end{pf}

\begin{Remark}
The choice of $\alpha$ determines the convergence speed of the estimators,
where a larger $\alpha$ enforces faster convergence of the estimates. 
\hfill $\Box$
\end{Remark}

\medskip

\begin{Remark}
In the case when $u_k \not\equiv 0$, the control inputs also have to be
added to the estimator dynamics, i.e. equation (\ref{eq:estimator}) has to
be modified as follows 
\begin{equation*}
\begin{aligned}
\dot{\hat{x}}^{(k)} =& A\ik \hat{x}^{(k)}+ B\ik u\ik + L\ik(z\ik-C\ik \hat{x}\ik) \\
& +K\ik \sum_{j=1}^N a_{kj}(M_j\ik \hat{x}_j^{(j)}- N_j\ik \hat{x}\ik),
\end{aligned}
\end{equation*}
where 
\begin{equation*}
B\ik = \begin{bmatrix}
B_k & 0 & 0 & 0 \\
0 & B_{j_1} & \hdots & 0 \\
0 & 0 & \ddots & 0 \\
0 & 0 & 0 & B_{j_{p_k}}
\end{bmatrix}, u\ik = \begin{bmatrix}
u_k \\ u_{j1} \\ \vdots \\ u_{j_{p_k}}
\end{bmatrix}.
\end{equation*}
\hfill $\Box$
\end{Remark}

As it can be seen from the LMIs \eqref{LMI}, the design problem has to be
solved in a centralized manner. However, this calculation is only done
a priori. The resulting cooperative estimators \eqref{eq:estimator} are 
local and their complexity only increases with the number of neighbors, not
with the total size of the network. 
In contrast, a direct application of the algorithms developed in
\citep{OlfatiSaber:2005vm}, \citep{Olfati-saber2006} and
\citep{Ugrinovskii2011}, \citep{Ugrinovskii2011a}, to the problem
considered here would result in the order of the estimators growing with
the size of the network. Therefore, the method presented in this paper is
scalable and guarantees $\mathcal{H}_\infty$ performance. 

\subsection{Detectability conditions}
Feasibility of the LMIs \eqref{LMI} is a sufficient condition for Problem 1 to have a solution, but it is not a necessary condition. However, simulations show that infeasibilty of \eqref{LMI} often relates to detectability issues.
Clearly, in order for the estimation algorithm to work, the global system
\eqref{sys:global} with the output matrix \eqref{sys:C_global} has to be
detectable. Moreover, as a stricter necessary condition, we have the
following proposition: 
\begin{prop}
Let $\widetilde{\mc{G}}$ be an iSCC of $\mc{G}$ and let the vertices in
$\widetilde{\mc{G}}$ be $v_1,...,v_\rho$, $\rho \leq N$, without loss of
generality. Let $\widetilde{E}$ be the component incidence matrix, which
only considers the edge set $\widetilde{\mc{E}}$. 

Then detectability of the system
\begin{equation}\label{sys:global_comp}
\begin{aligned}
\dot{\widetilde{x}}=& \begin{bmatrix}
A_1 & & \\
& \ddots & \\
& & A_\rho
\end{bmatrix} \widetilde{x} \\
y =& \left( \widetilde{E} \otimes I_r \right)
\begin{bmatrix}
C_1 & & \\
& \ddots & \\
& & C_\rho
\end{bmatrix} \widetilde{x}, 
\end{aligned}
\end{equation}
with $\widetilde{x} = [x_1^\top,...,x_\rho^\top]^\top$ is a necessary condition for the existence of a solution to Problem 1.
\end{prop}
\begin{pf}
This proposition follows from the fact that the agents lying in $\widetilde{\mathcal{G}}$ do not receive any information from outside $\widetilde{\mathcal{G}}$.
\flushright$\blacksquare$ 
\end{pf}

\begin{Remark}
Detectability of the iSCC can be well analyzed using the theory presented
in \cite{Zelazo2008}.
\end{Remark}

\section{Synchronization using relative imperfect measurements}

The cooperative estimator design introduced above can be combined with
various synchronization methods. In this section we study the combination
of a synchronization algorithm presented in \citep{Wieland2011} with the
observer design from Section 3. This results in a synchronization scheme in
which local controllers rely on state estimates obtained from imperfect
relative output information. We also analyze disturbance attenuation
properties of such controllers. 

The controller is proposed in the following form:
\begin{align}\label{eq:zeta}
\dot{\zeta}_k &= S \zeta_k + \sum_{j=1}^N a_{kj}(\zeta_j - \zeta_k) \\ \label{eq:u_k}
u_k &= \Lambda_k \zeta_k + H_k(\hat{x}\ik_k-\Pi_k \zeta_k),
\end{align}
where the matrices $S \in \mathbb{R}^{\nu \times \nu}, R \in \mathbb{R}^{r
  \times \nu}, \Pi_k \in \mathbb{R}^{n_k \times \nu}$ and $\Lambda_k \in
\mathbb{R}^{m_k \times \nu}$ are obtained from the Francis equations
\begin{equation}\label{eq:francis}
\begin{aligned}
A_k \Pi_k + B_k \Lambda_k &= \Pi_k S  \\
C_k \Pi_k &= \Gamma,
\end{aligned}
\end{equation}
where $k=1,...,N$ and $\nu$ is a positive integer which determines the
dimension of the internal model for each controller. It is
shown in \citep{Wieland2010b} that under mild assumptions on the closed
loop systems, solvability of \eqref{eq:francis} with $(S,\Gamma)$ being
observable is a necessary condition for synchronization of the unperturbed
systems \eqref{eq:LTI}. Therefore the assumption of the existence of a
solution to \eqref{eq:francis} is not a conservative assumption.

In the following, the term $\epsilon_k = x_k - \Pi_k \zeta_k$ will be
called the local regulation error. As this error is the quantity which
\eqref{eq:u_k} regulates to zero, our performance analysis will focus on
this characteristic of the system.

\textbf{Problem 2:}
Find $H_k$, $k=1,...,N$, such that the following two properties are
satisfied simultaneously. 
\begin{compactenum}[(i)]
\item Exact synchronization is achieved in the absence of disturbance and
  measurement disturbances, i.e. when $\xi_k=0$, $\eta=0$, then $y_j
  -y_k \to 0$ for all $j,k=1,...,N$. 

\item $\mathcal{H}_{\infty}$ performance with respect to the regulation
  error is achieved in the sense that 
\begin{equation}\label{eq:hinf-performance-sync}
\begin{aligned}
& \sum_{k=1}^N \int_0^\infty \epsilon_k^\top R_k \epsilon_k dt
\leq  \sum_{k=1}^N \int_0^\infty (\kappa^2 \| \xi_k \|^2 + \theta^2 \| \eta\ik \|^2)dt \\
& \!+\!\! \sum_{k=1}^N \epsilon_{k,0}^\top X_k \epsilon_{k,0} \!+\!\! I_0 \!+\!\! \mu^2 \sum_{k=1}^N \int_0^\infty \!\!\! \| \sum_{j=1}^N a_{kj} (\zeta_j - \zeta_k) \|^2 dt
\end{aligned}
\end{equation}
with positive definite weighting matrices $R_k$ and $X_k$, for $k=1,...,N$,
and real parameters $\mu, \kappa, \theta >0$ to be defined later. 
\end{compactenum}

The last term of \eqref{eq:hinf-performance-sync}  
characterizes the effect
of the uncertainty that the controller for agent $k$ has in relation to the
initial states of the internal models of its neighbours. This term has to be
included into the performance inequality because the internal system
\eqref{eq:zeta} is neither affected by the estimators, nor by the controllers
\eqref{eq:u_k}. Therefore, large initial mismatches between $\zeta_k$,
$k=1,...,N$ will generate significant transient dynamics which will affect
output regulation error and controller performance. In the case when
$\zeta_k$, $k=1,...,N$, are initialized identically, this term vanishes. 

\begin{thm}\label{thm:sync}
Consider an interconnected system consisting of $N$ LTI systems
\eqref{eq:LTI}, with the communication topology represented by a connected graph
$\mathcal{G}=\{\mathcal{V}, \mathcal{E}, \mathcal{A}\}$ and the dynamic
local controllers given as \eqref{eq:zeta}, \eqref{eq:u_k}. 
Suppose the following conditions hold:
\begin{enumerate}
\item The Francis equations \eqref{eq:francis} have a solution $(S,\Gamma)$
  which is an observable pair, with $\sigma(S) \subset j\mathbb{R}$. 

\item There exist constants $\mu>0$ and $\lambda>0$ such that for each $k=1,\ldots, N$, the
  algebraic Riccati equation
\begin{equation}\label{eq:ricatti}
\begin{aligned}
&X_k A_k \!+\!\! A_k^\top \! X_k + R_k \\
&- X_k \left(\frac{1}{\lambda^2} B_k B_k^\top \!\!-\!\! \frac{1}{\mu^2}\widetilde{B}_k \widetilde{B}_k^\top \!\!-\! \frac{1}{\mu^2}\Pi_k \Pi_k^\top \right) X_k
 = 0
\end{aligned}
\end{equation}
has a positive definite symmetric solution $X_k$ such that
$A_k-\frac{1}{\lambda^2}B_kB_k'X_k$ is a Hurwitz matrix. 

\item For a given $\gamma>0$, Problem 1 with the weighting
  matrices
  \begin{equation}\label{eq:performance_weighting}
W\ik=\begin{bmatrix}
 \frac{1}{\lambda^2} X_kB_k B_k^\top X_k & 0 \\
0 & 0 \\
\end{bmatrix}
\end{equation}
has a solution. 
\end{enumerate}
Then the collection of matrices $H_k$ defined as
\begin{equation}\label{eq:Kk}
H_k =-\frac{1}{\lambda^2} B_k^\top X_k
\end{equation}
solves Problem 2.

\end{thm}

\begin{pf} Suppose the conditions of Theorem \ref{thm:sync} hold. Note that the group of controller states $\zeta_k$ does
  not depend on the system states $x_k$ and estimator states $x
  \ik_k$. Using Lemma 1 from \citep{Scardovi2009}, we know that
  $\zeta_j-\zeta_k \to 0$ exponentially for all $k,j = 1,...,N$. 

Now, we consider the error variable $\epsilon_k = x_k - \Pi_k
\zeta_k$. Using equations \eqref{eq:zeta} and \eqref{eq:u_k} the derivative
of $\epsilon_k$ is found to be
\begin{align*}
\dot{\epsilon}_k =& \dot{x}_k - \Pi_k \dot{\zeta}_k \\
=& A_k x_k + B_k \Lambda_k \zeta_k + B_k H_k(\hat{x}\ik_k-\Pi_k \zeta_k) +\widetilde{B}_k \xi_k \\
&- \Pi_k S \zeta_k - \Pi_k\sum_{j=1}^N a_{kj}(\zeta_j - \zeta_k).
\end{align*}
It follows from \eqref{eq:francis} that $B_k \Lambda_k \zeta_k - \Pi_k S
\zeta_k = - A_k \Pi_k \zeta_k$ and therefore 
\begin{align*}
\dot{\epsilon}_k =& A_k x_k - A_k \Pi_k \zeta_k + B_k H_k(x_k-\Pi_k \zeta_k) +\widetilde{B}_k \xi_k \\
& - \Pi_k\sum_{j=1}^N a_{kj}(\zeta_j - \zeta_k) - B_k H_k (x_k - \hat{x}\ik_k) \\
=& (A_k +B_kH_k) \epsilon_k +\widetilde{B}_k \xi_k \\
& - \Pi_k\sum_{j=1}^N a_{kj}(\zeta_j - \zeta_k) - B_k H_k e\ik_k.
\end{align*}
Now, we consider the Lyapunov function
\begin{equation}\label{eq:lyap2}
V(\epsilon)= \sum_{k=1}^N \underbrace{\epsilon_k^\top X_k \epsilon_k}_{V_k(\epsilon_k)},
\end{equation}
where $X_k$ is the solution to \eqref{eq:ricatti}. With \eqref{eq:Kk} and
\eqref{eq:ricatti}, an upper bound on the Lie derivative of
$V_k(\epsilon_k)$ can be obtained using the completion of squares argument:  

\begin{align*}
\dot{V}_k =& \epsilon_k^\top \left( A_k^\top X_k + X_k A + 2X_kB_kH_k \right) \epsilon_k \\
&+ 2\epsilon_k^\top \! X_k \left( \! \widetilde{B}_k \xi_k - \Pi_k \sum_{j=1}^N a_{kj} (\zeta_j - \zeta_k) - B_k H_k e_k \! \right) \\
\leq & \epsilon_k^\top \left( A_k^\top X_k + X_k A + 2X_kB_kH_k \right) \epsilon_k \\
&+ \epsilon_k^\top \! X_k \left( \! \frac{1}{\mu^2} \widetilde{B}_k \widetilde{B}_k^\top  + \frac{1}{\mu^2} \Pi_k \Pi_k^\top + \frac{1}{\lambda^2}B_k B_k ^\top \! \right) X_k \epsilon_k \\
&+ \mu^2 \left( \| \xi_k \|^2 + \| \sum_{j=1}^N a_{kj} (\zeta_j - \zeta_k)
  \|^2\right) + \| e_k \|^2_{\lambda^2 H_k ^\top H_k}  \\
=& \epsilon_k^\top \left( A_k^\top X_k + X_k A 
+ \frac{1}{\mu^2}X_k \left( \widetilde{B}_k \widetilde{B}_k^\top  + \Pi_k
   \Pi_k^\top \right) X_k\right)\epsilon_k \\
 &+ \epsilon_k^\top \left(2X_kB_kH_k + 
\frac{1}{\lambda^2} X_k B_k B_k ^\top X_k\right) \epsilon_k \\
&+ \mu^2 \left( \| \xi_k \|^2 + \| \sum_{j=1}^N a_{kj} (\zeta_j - \zeta_k)
   \|^2\right) + \| e_k \|^2_{\lambda^2 H_k ^\top H_k} \\
=& \epsilon_k^\top \left( A_k^\top X_k + X_k A 
+ \frac{1}{\mu^2}X_k \left( \widetilde{B}_k \widetilde{B}_k^\top  + \Pi_k
   \Pi_k^\top \right) X_k\right)\epsilon_k \\
 &+ \epsilon_k^\top (\lambda H_k+\frac{1}{\lambda}B_k^\top X_k)^\top (\lambda H_k+\frac{1}{\lambda}B_k^\top X_k) \epsilon_k \\ 
& - \lambda^2 \epsilon_k^\top H_k^\top H_k \epsilon_k \\
&+ \mu^2 \left( \| \xi_k \|^2 + \| \sum_{j=1}^N a_{kj} (\zeta_j - \zeta_k)
   \|^2\right) + \| e_k \|^2_{\lambda^2 H_k ^\top H_k} 
\end{align*}
Now let $H_k$ be defined as in \eqref{eq:Kk}. Subsequently, it holds that
 \begin{align*}
\dot{V}_k \le& 
 \epsilon_k^\top \left( A_k^\top X_k + X_k A 
+ \frac{1}{\mu^2}X_k \left( \widetilde{B}_k \widetilde{B}_k^\top  + \Pi_k
   \Pi_k^\top \right) X_k\right)\epsilon_k \\
& - \lambda^2 \epsilon_k^\top H_k^\top H_k \epsilon_k \\
&+ \mu^2 \left( \| \xi_k \|^2 + \| \sum_{j=1}^N a_{kj} (\zeta_j - \zeta_k)
   \|^2\right) + \| e_k \|^2_{\lambda^2 H_k ^\top H_k} .
\end{align*}
Using the Riccati equation \eqref{eq:ricatti} we finally obtain
 \begin{align*}
\dot{V}_k \le& 
- \epsilon_k^\top R_k\epsilon_k \\
&+ \mu^2 \left( \| \xi_k \|^2 + \| \sum_{j=1}^N a_{kj} (\zeta_j - \zeta_k)
   \|^2\right) + \| e_k \|^2_{ \lambda^2H_k ^\top H_k}.
\end{align*}

In the absence of perturbation, we know that $\zeta_j -
\zeta_k \to 0$ and $x_k - \hat{x}\ik_k \to 0$ exponentially fast for all $k,j=1,...,0$. Since $A-\frac{1}{\lambda^2}B_kB_k^\top X_k$ is
Hurwitz, we can now conclude that $\epsilon_k \to 0$ for all $k=1,...,N$, i.e. Property (i) of
Problem 2 is satisfied. 

On the other hand, when the system is affected by disturbances, integrating 
the above inequality over the interval $[0, T]$ leads to the following inequality 
\begin{align*}
V(\epsilon & (T)) +\sum_{k=1}^N \int_0^T \epsilon_k^\top R_k \epsilon_k dt\\
\leq \; & \mu^2  \sum_{k=1}^N \int_0^T \| \xi_k \|^2 dt + \mu^2 \sum_{k=1}^N \int_0^T \| \sum_{j=1}^N a_{kj} (\zeta_j - \zeta_k) \|^2 dt \\ 
&+  \sum_{k=1}^N \int_0^T \| e_k \|^2_{\lambda^2 H_k^\top H_k} + \sum_{k=1}^N \epsilon_{k,0}^\top X_k \epsilon_{k,0}.
\end{align*}
Using the definition of the weight $W\ik$ \eqref{eq:performance_weighting} and the performance property of the distributed estimator established
in Property (ii) of Problem 1, we can conclude
\begin{align*}
\sum_{k=1}^N & \int_0^T \epsilon_k^\top R_k \epsilon_k dt\\
\leq \; & \mu^2  \sum_{k=1}^N \int_0^T \| \xi_k \|^2 dt + \mu^2 \sum_{k=1}^N \int_0^T \| \sum_{j=1}^N a_{kj} (\zeta_j - \zeta_k) \|^2 dt \\ 
&\!+\! \gamma^2  \sum_{k=1}^N \int_0^T (\| \xi\ik \|^2 + \| \eta\ik \|^2)dt + I_0 + \sum_{k=1}^N \epsilon_{k,0}^\top X_k \epsilon_{k,0} \\
\leq \; & \kappa^2 \sum_{k=1}^N \int_0^T  \| \xi_k \|^2 dt + \mu^2 \sum_{k=1}^N \int_0^T  \| \sum_{j=1}^N a_{kj} (\zeta_j - \zeta_k) \|^2 dt \\ 
&+ \theta^2 \sum_{k=1}^N \int_0^T \| \eta\ik \|^2 dt + I_0 + \sum_{k=1}^N \epsilon_{k,0}^\top X_k \epsilon_{k,0},
\end{align*}
where $\kappa^2 = \mu^2 + q_{max}\gamma^2$ and $\theta^2 = \gamma^2$. Letting $T\to \infty$ shows that Property (ii) of Problem 2 holds.
\flushright $\blacksquare$
\end{pf}

Performance bound \eqref{eq:hinf-performance-sync} may be conservative
since in this paper we did not optimize performance of the
closed loop system. Improving this bound as well as analysis of other
synchronization control schemes will be the subject of our future research. 

\section{Simulation Example}
\begin{figure}
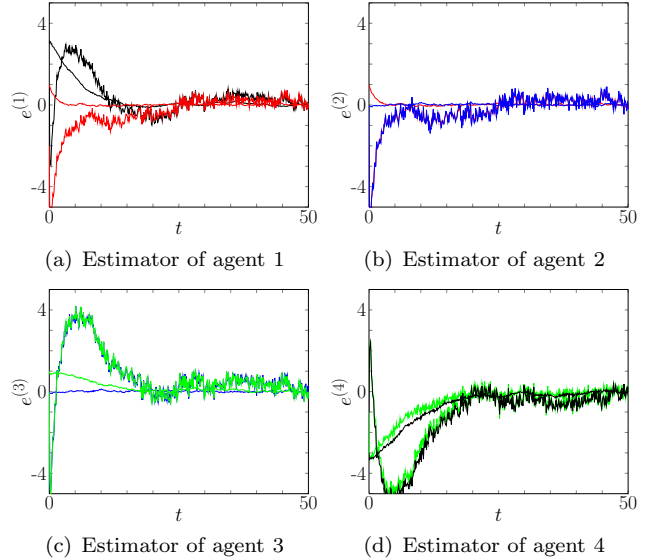
 \label{fig:estimation_error}
 \centering
 	\subfigure[Estimator of agent $1$]
 	{\input{error1.tex}}
 		\subfigure[Estimator of agent $2$]
 	{\input{error2.tex}}
 		\subfigure[Estimator of agent $3$]
 	{\input{error3.tex}}
 		\subfigure[Estimator of agent $4$]
 	{\input{error4.tex}}
	\caption{Plots of the local estimation errors. Black lines represent the estimation error of agent $1$. Red, blue, and green lines represent the estimation error of agent $2, 3$ and $4$, respectively.}	
\end{figure}

To illustrate the cooperative estimators \eqref{eq:estimator} we take four agents connected in a cyclic topology, corresponding to Figure \ref{fig:estimator_setup}. The four agents' dynamics are
described by \eqref{eq:LTI}, where the matrices are 
\begin{equation*}
\begin{aligned}
A_1=&\begin{bmatrix}
\; 0 \; & \;\; 1 \; \\ \; 0 \; & \;\; 0 \;
\end{bmatrix} &
A_2=&\begin{bmatrix}
0 & 1 \\ 0 & -1
\end{bmatrix} \\
A_3=&\begin{bmatrix}
0.1 & 1 \\ 0 & -1
\end{bmatrix} &
A_4=&\begin{bmatrix}
0.1 & 1 \\ 0 & 0
\end{bmatrix}.
\end{aligned}
\end{equation*}
The input and output matrices are given as
\begin{equation*}
\begin{aligned}
B_k=\begin{bmatrix}
0 \\ 1
\end{bmatrix} &&
\widetilde{B}_k=\begin{bmatrix}
0 \\ 0.5
\end{bmatrix} &&
C_k =\begin{bmatrix}
1 & 0
\end{bmatrix}  && \omega=0.1
\end{aligned}
\end{equation*}
for all $k=1,2,3,4$. 
Note that for all $k$, the pair $(A\ik, C\ik)$ is not detectable.
When we consider the estimation problem only, i.e. we design the local estimators \eqref{eq:estimator} by solving the LMIs \eqref{LMI}, we achieve a performance bound of $\gamma = 5.61$. The parameters are chosen as
\begin{equation*}
W_k=\begin{bmatrix}
I_2 & 0 \\ 0 & 0
\end{bmatrix} \quad \alpha=0.1 \quad \pi_k = 0.025.
\end{equation*}
Simulations of the local estimators are shown in Figure \ref{fig:estimation_error}.

To put the performance into perspective: The centralized $H_\infty$-estimator for system \eqref{sys:global} with output \eqref{sys:C_global} achieves a performance of $\gamma=5.19$.

Next, we consider the synchronization problem. For the given systems, the Francis Equations \eqref{eq:francis} are solvable with the matrices
\begin{align*}
&& S&=\begin{bmatrix}
0 & 0 \\ 0 & 1
\end{bmatrix} &\!\!\! \Gamma &=\begin{bmatrix}
1 & 0
\end{bmatrix} \\
\Pi_1&=\begin{bmatrix}
1 & 0 \\ 0 & 1
\end{bmatrix} &\!\!\!  \Pi_2&=\begin{bmatrix}
1 & 0 \\ 0 & 1
\end{bmatrix} &\!\!\! \Pi_3&= \begin{bmatrix}
1 & 0 \\ -0.1 & 1
\end{bmatrix}  &\!\!\! \Pi_4&= \begin{bmatrix}
1 & 0 \\ -0.1 & 1
\end{bmatrix} \\
\Lambda_1&= \begin{bmatrix}
0 & 0 
\end{bmatrix}
 &\!\!\! \Lambda_2&= \begin{bmatrix}
0 & 1 
\end{bmatrix}
&\!\!\! \Lambda_3&= \begin{bmatrix}
-0.1 & 0.9
\end{bmatrix}
 &\!\!\! \Lambda_4&= \begin{bmatrix}
0 & -0.1
\end{bmatrix}.
\end{align*}

The algebraic Ricatti-equation \eqref{eq:ricatti} with $R_k=I_2$ for $k=1,2,3,4$ has a solution for $\mu=1.2$ and $\lambda=0.1$. Then, we take the resulting weights $W_k$ given by \eqref{eq:performance_weighting} and solve the LMIs \eqref{LMI} for the estimators. Using this design method, we achieve performance bounds of $\kappa=11.50$ and $\theta=11.44$.

\section{Conclusion}

In this paper, we presented an LMI-based solution to the design problem of local estimators, which estimate local state variables within an heterogeneous multi-agent system. The dimension of the local estimators do not grow with the number of agents in the network. In particular, every local estimator may deal with a local system that is even not detectable, as cooperation between the local estimators establishes detectability. Moreover, the presented algorithm guarantees $\mathcal{H}_\infty$-performance for both the estimators and for synchronization. Further work will include the application to other synchronization algorithms, time-varying topologies and separate measurement and communication topologies.

\bibliography{ifacconf}   


\end{document}